# The mixed-state Hall conductivity of single-crystal films $Nd_{2-x}Ce_xCuO_{4+\delta}$ (x=0.14).

Shelushinina N.G.[1], Harus G.I.[1], Charikova T.B.[1,2], Petukhov D.S.[1], Petukhova O.E.[1], Ivanov A.A.[3]

[1]M.N. Mikheev Institute of Metal Physics Ural Branch of RAS, Ekaterinburg, Russia,
[2]Ural Federal University, Ekaterinburg, Russia,
[3]National Research Nuclear University MEPhI, Moscow, Russia

The magnetic-field dependencies of the longitudinal and Hall resistivity of the electron-doped compounds $Nd_{2-x}Ce_xCuO_{4+\delta}$ in underdoped region ($x$=0.14) were investigated. It was established experimentally a strong magnetic field dependence of the Hall conductivity, $\sigma_{xy}(B)=C-b/B$, in the region of magnetic fields corresponding to a transition from superconducting to resistive state. The observed feature can be explained with the sum of contributions of the quasiparticles and moving Abrikosov vortices into Hall effect in a mixed state of type-II superconductor.

Introduction

One of the most striking features of the vortex motion in the oxide superconductors is the behavior of the Hall resistivity which attracted imperishable attention.

In a type-II superconductor in an applied magnetic field $B$, quantized magnetic vortices (fluxes) are formed by supercurrents, at the mixed state, for $B$ larger than the lower critical field $B_{c1}$ and less than the upper critical field $B_{c2}$. The high-temperature superconductors exhibit puzzling Hall effect phenomena in the mixed state. One of the most puzzling of these features is a reversal of the sign of the Hall effect in the neighborhood of the superconducting transition as the temperature or the magnetic field is varied [1].

The Hall sign is determined by the topology of the Fermi surface in the normal state, while it is determined by the vortex motion in the superconducting state. The classical theories of vortex motion, the Bardeen-Stephen [2] and Nozieres-Vinen [3] models, predict that the superconducting and normal states will have the same Hall sign, and thus cannot explain this anomaly.

A number of theoretical predictions have been made concerning the behavior of the flux-flow electric transport coefficients in type-II superconductors. Several attempts to understand the Hall anomaly have been undertaken, but the microscopic origin of this phenomenon remains a controversial [4].

A phenomenological theory based on the time dependent Ginzburg-Landau (TDGL) equation has been shown to be quite successful in describing the Hall effect in the superconducting state [5, 6]. According to the TDGL theory, the vortex Hall conductivity arising from the hydrodynamic contribution plays an important role in determining the Hall sign at low fields.

In a light of recent theoretical developments it seems useful to further analyze the resistivity and Hall-effect data in a mixed state of diverse materials. The systematic investigations of magnetic-field dependence of longitudinal and Hall resistances in the electron-doped compounds $Nd_{2-x}Ce_xCuO_{4+\delta}$ at underdoped region ($x$=0.14) with varying degrees of disorder ($\delta$) were held by us. In our previous work [7] the correlation between the longitudinal electrical resistivity and the Hall resistivity in the region of magnetic fields corresponding to a transition from superconducting to the normal states was established and has been analyzed on the basis of scaling relations.

The aim of our present study is to find out features of the Hall effect in the vortex state for the disordered $Nd_{2-x}Ce_xCuO_{4+\delta}$ system at the under doped region ($x$=0.14) which is on the border of the antiferromagnetic and superconducting phases in electron-doped cuprates [8].

**Experimental results and discussion**

The resistivity and Hall-effect data were obtained simultaneously on a sample of electron-doped high-temperature superconductor $Nd_{2-x}Ce_xCuO_{4+\delta}$ ($x$=0.14), with the applied magnetic field $B$ oriented perpendicular to the copper-oxygen planes (z-direction). The current density $j$ in the sample was in the x-direction, and the Hall electric field $E_H$ was in the $y$-direction, indicating negatively charged current carriers (electrons) in the normal state. The measurements of the longitudinal resistivity $\rho_{xx}$ and Hall resistivity $\rho_{xy}$ as functions of the external magnetic field $B$ up to 12T in the temperature range $T$=(0.53-40)K in a mixed and a normal states were made. The data for single crystal films $Nd_{2-x}Ce_xCuO_{4+\delta}$ with disorder parameter $k_Fl$=6.0 are presented on the fig. 1 (the parameter $k_Fl$ which serves as a measure of disorder in a system was found from the experimental value of $\rho_{xx}$ [7]).

Because Hall conductivity is typically defined as $\sigma_{xy}=\rho_{xy}/\rho^2_{xx}$ (by assuming $\rho_{xx}<<\rho_{xy}$) it is convenient to discuss the Hall results using $\sigma_{xy}(B)$ [9]. In fig. 2 the field dependences of the Hall conductivity for $Nd_{2-x}Ce_xCuO_{4+\delta}$ film ($k_Fl$=6.0) are shown for various temperatures. It can be seen from fig. 2 that in the immediate vicinity of the transition from the superconducting to the resistive state the Hall conductivity tends to diverge to a large positive (at $T$=0.53K) or negative (at $T \geq 0.7K$) value with decreasing field.

A more detailed picture for $\sigma_{xy}(B)$ dependencies in the mixed state is presented on fig. 3. The data for $T$=0.53K with a reversal of the sign of the Hall effect from negative in the normal state ($B$>7.4T) to positive in the mixed state ($B$<7.4T) are given in a separate drawing on the inset of fig. 3 where the comparison of $\sigma_{xy}(B)$ and $\rho_{xx}(B)$ dependences is also shown. It is seen that the field dependence of $\sigma_{xy}(B)$ changes approximately as $1/B$ at $T$=8-10K and much more rapidly at low $T$=0.53-1.35K.

The $\sigma_{xy}(B)$ dependence of such a type has repeatedly been observed in the mixed state of high-$T_c$ superconductors $Nd_{2-x}Ce_xCuO_{4+\delta}$ [10, 11] and $Sm_{2-x}Ce_xCuO_{4+\delta}$ [10], of $Tl_2Ba_2CaCu_2O_8$ thin films [11], of $Tl_2Ba_2CaCu_2O_8$ epitaxial film and $YBa_2Cu_3O_7$ single crystal before and after irradiation [12], of untwined single-crystal $YBa_2Cu_3O_{7-\delta}$ [13], of $La_{2-x}Sr_xCuO_4$ single-crystal thin films [14], of various $p$-type high-$T_c$ cuprates including La, Y-, Bi-based compounds [15], of Hg-based superconducting thin films [16], of high-quality $Bi_2Sr_2CuO_x$ single crystals [17], of $Ba(Fe_{1-x}Co_x)_2As_2$ epitaxial film [18] and of Fe(Te,S) single crystal [19].

The conventional explanation of such $\sigma_{xy}(B)$ behavior is based on a microscopic approach using the time-dependent Ginzburg-Landau theory which has been proposed by

Dorsey [5] and by Kopnin et al. [6]. According to this model, there are two contributions to the $\sigma_{xy}(B)$ in the mixed state:

$$\sigma_{xy}(B)=\sigma_{xy}^{n}(B)+\sigma_{xy}^{f}(B), \qquad (1)$$

where $\sigma_{xy}^{n}(B)$ is the conductivity of normal quasiparticles that experience a Lorentz force inside the vortex core. The second term $\sigma_{xy}^{f}(B)\sim 1/B$ is an anomalous contribution due to the motion of vortices parallel to the electrical current density *j*. As a possible origin of the longitudinal component of the vortex velocity, Kopnin et al. [6] considered the vortex-traction force by a transport supercurrent.

The quasiparticle term, $\sigma_{xy}^{n}$, has the same sign as it is in the normal state. Accordingly the sign reversal of the Hall effect can occur if the vortex term, $\sigma_{xy}^{f}$, has an opposite sign to $\sigma_{xy}^{n}$. However, the factors that fix the sign of $\sigma_{xy}^{f}$ are not clear. A few microscopic calculations have suggested that the vortex term can change its sign from the normal state depending on the detailed electronic structure of the material.

A more specific mechanism for the sign change of the Hall effect in the flux flow region is proposed by Feigel'man et al. [20]. The difference $\delta n$ between the electron density at the center of the vortex core and that far outside the vortex causes the additional contribution to the Hall conductivity $\sigma_{xy}^{f}=-e\delta n/B$. This contribution can be larger than the conventional one in the dirty case $\Delta(T)\tau<1$. If the carrier density inside the core exceeds that far outside, a sign change may occur as a function of temperature.

Note that in the model of ref. [20] the observed in our system quite strong dependence of $\sigma_{xy}^{f}$ on the temperature (until the sign change) as well as the stronger than *1/B* dependence on magnetic field may well be explained by the $\delta n$ dependencies both on *T* and *B*, $\delta n(B, T)$.

At low magnetic field the $\sigma_{xy}^{f}$ is the dominant term but at higher field $\sigma_{xy}^{n}$ are important and should dominate over $\sigma_{xy}^{f}(B)$. If $\sigma_{xy}^{f}(B)$ has a different sign when compared to $\sigma_{xy}^{n}(B)$, it is possible to observe a sign reversal in the Hall effect in the superconducting state (see, for example, our data at 0.53K).

**Conclusions**

The magnetic-field dependencies of longitudinal and Hall resistivity were investigated in the electron-doped compound $Nd_{2-x}Ce_xCuO_{4+\delta}$ at underdoped region (*x*=0.14) which is on the border of the antiferromagnetic and superconducting phases in electron-doped cuprates. The special attention was given to features of the Hall effect in the mixed state of this superconducting material.

We have established that in a vicinity of the transition from the superconducting to the resistive state the Hall conductivity tends to diverge to a large positive or negative value with decreasing field in analogy with the $\sigma_{xy}(B)$ dependencies repeatedly observed in the mixed state both of *p*-type oxide superconductors and of Fe-based ones.

Within the framework of current theoretical concepts such a behavior of $\sigma_{xy}(B)$ arises from an anomalous contribution $\sigma_{xy}^{f}(B)) \sim 1/B$ due to the motion of vortices parallel to the electrical current density in a flux-flow regime. It seems important to demonstrate the versatility of this contribution in a mixed state of manifold superconducting systems.

Referencies

1. S. J. Hagen *et al.,* Phys. Rev. B **47**, 1064 (1993).

2. J. Bardeen and M. J. Stephen, Phys. Rev. **140**, A1197(1965).

3. P. Nozieres and W. F. Vinen, Philos. Mag. **14**, 667 (1966).

4. G. Blatter, M. Y. Feigel'man, Y. B. Geshkenbein, A. I. Larkin, V. M. Vinokur, Rev. Mod. Phys., **66**, 1I25 (1994)

5. Alan T. Dorsey, Phys. Rev. 8 46, 8376 (1992).

6. N. B. Kopnin, B.I. Ivlev, and V. A. Kalatsky, J. Low Temp. Phys, **90**, 1 (1993); N.B. Kopnin, A.V. Lopatin, Phys. Rev. B 51 (1995) 15291.

7. T.B.Charikova, N.G.Shelushinina, G.I.Harus, D.S.Petukhov, O.E. Petukhova, A.A.Ivanov. Physica C **525–526**, 78–83 (2016).

8. N. P. Armitage, P. Fournier, R. L. Greene, Rev.Mod.Phys. 82, 2421 (2010).

9. V.M.Vinokur, V.B.Geshkenbein, M.V. Feigel'man, G. Blatter, Phys.Rev.Lett. **71**, 1242 (1993).

10. M.Cagigal, J.Fortcuberta, M.A.Crusellas et al., Physica C 248, 155 (1995).

11. T. W. Clinton, A. W. Smith, Qi Li, J. L. Peng, R. L. Greene, and C. J. Lobb, M. Eddy, C. C. Tsuei, Phys.Rev.B 52, R7046 (1995)

12. A. V. Samoilov, Z. G. Ivanov, and L.-G. Johansson, Phys.Rev. B 49, 3667 (1994)

13. D. M. Ginsberg and J. T. Manson, Phys. Rev. B 51, 515 (1995). YBaCuO

14. Y. Matsuda, T. Nagaoka, G. Suzuki, K. Kumagai, Minoru Suzuki, M. Machida, M. Sera, M. Hiroi, and N. Kobayashi, Phys. Rev. B 52, 15749 (1995).LCCO

15. T. Nagaoka, Y. Matsuda, H. Obara, A. Sawa, T. Terashima, I.Chong, M. Takano, and M. Suzuki,Phys. Rev. Lett. **80**, 3594 (1998).

16. W.-S. Kim, W. N. Kang, S. J. Oh, M.-S. Kim, Y. B., S.-I. Lee, C. H. Choi and H.-C. Ri, cond-mat 9904385.

17. S. I. Vedeneev, A. G. M. Jansen, and P. Wyder, JETP, 90, 1042 (2000).

18. Hikaru Sato,Takayoshi Katase,Won Nam Kang, HidenoriHiramatsu,Toshio Kamiya,and Hideo Hosono, Phys.Rev.B **87**, 064504 (2013)

19. Hechang Lei, Rongwei Hu,E. S. Choi, and C. Petrovic, Phys.Rev.B**82**, 134525 (2010)

20. M. V. Feigel'man, V. B. Geshkenbein, A. I. Larkin, and V. M. Vinokur, cond-mat/9503082

Figure captions

Figure 1. Magnetic field dependencies of the longitudinal resistivity $\rho_{xx}(B)$ and Hall resistivity $\rho_{xy}(B)$ for single crystal films of underdoped ($x=0.14$) $Nd_{2-x}Ce_xCuO_{4+\delta}$ with disorder parameters $k_Fl= 6.0$ at different temperatures.

Figure 2. The magnetic-field dependences of the Hall conductivity, $\sigma_{xy}(B)$, for $Nd_{2-x}Ce_xCuO_{4+\delta}$ film in the underdoped region ($x=0.14$) for various temperatures.

Figure 3. The detailed picture for $\sigma_{xy}(B)$ dependencies in the mixed state of $Nd_{2-x}Ce_xCuO_{4+\delta}$ film for various temperatures. Inset shows a comparison of $\sigma_{xy}(B)$ and $\rho_{xx}(B)$ dependences at $T=0.53K$.

Figure 1

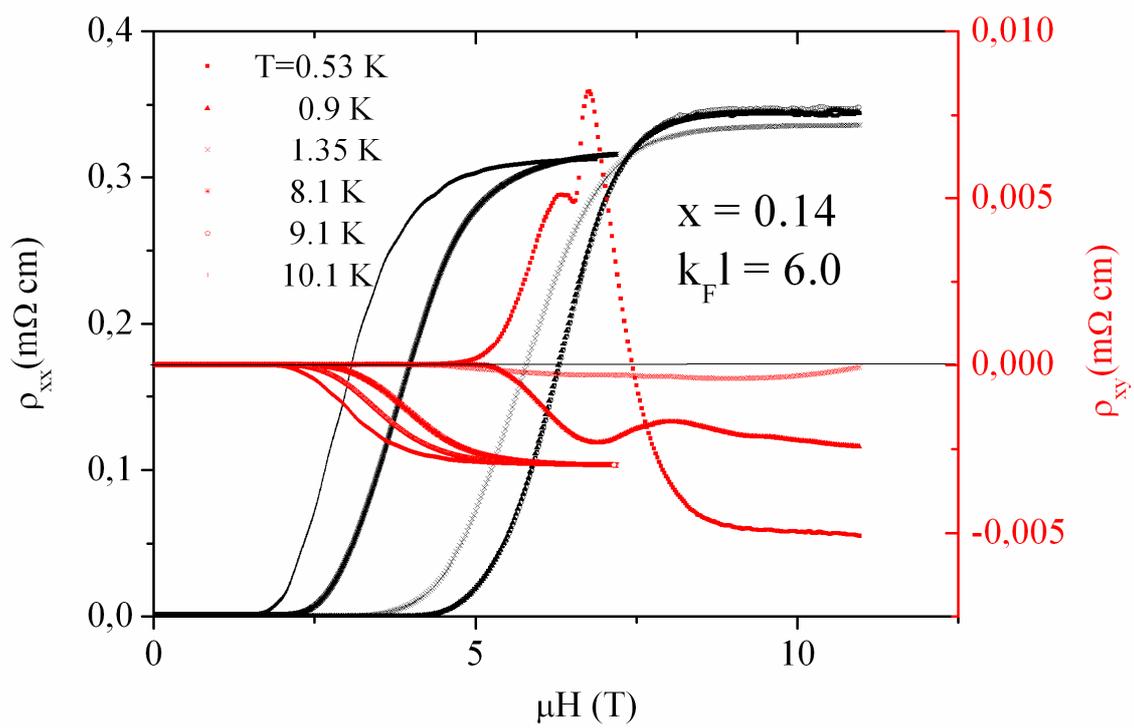

Figure 2

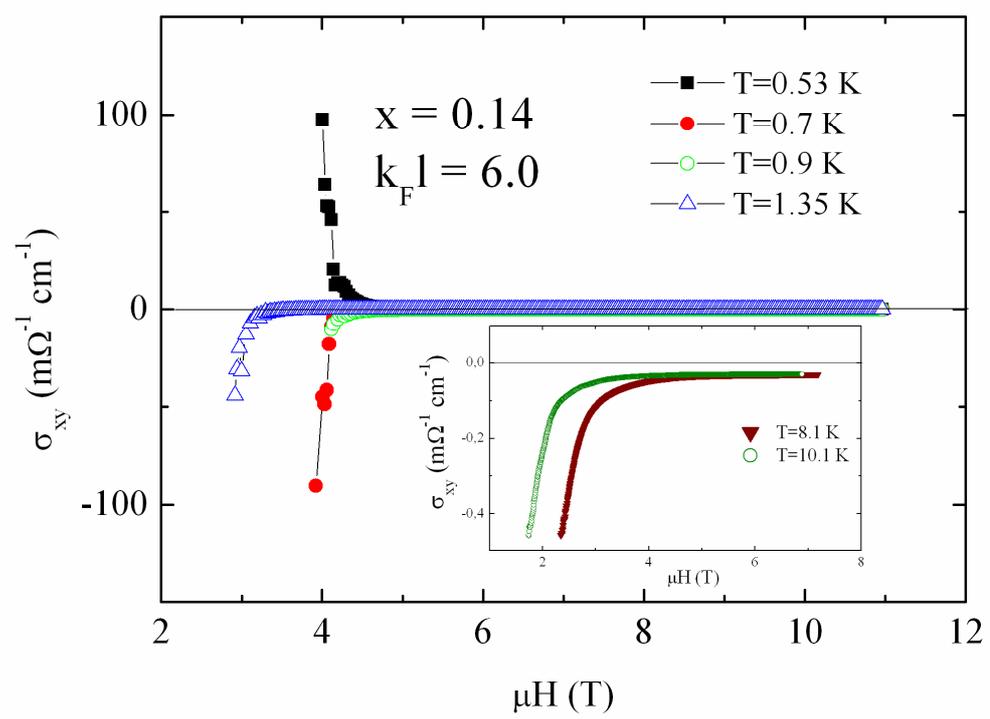

Figure 3

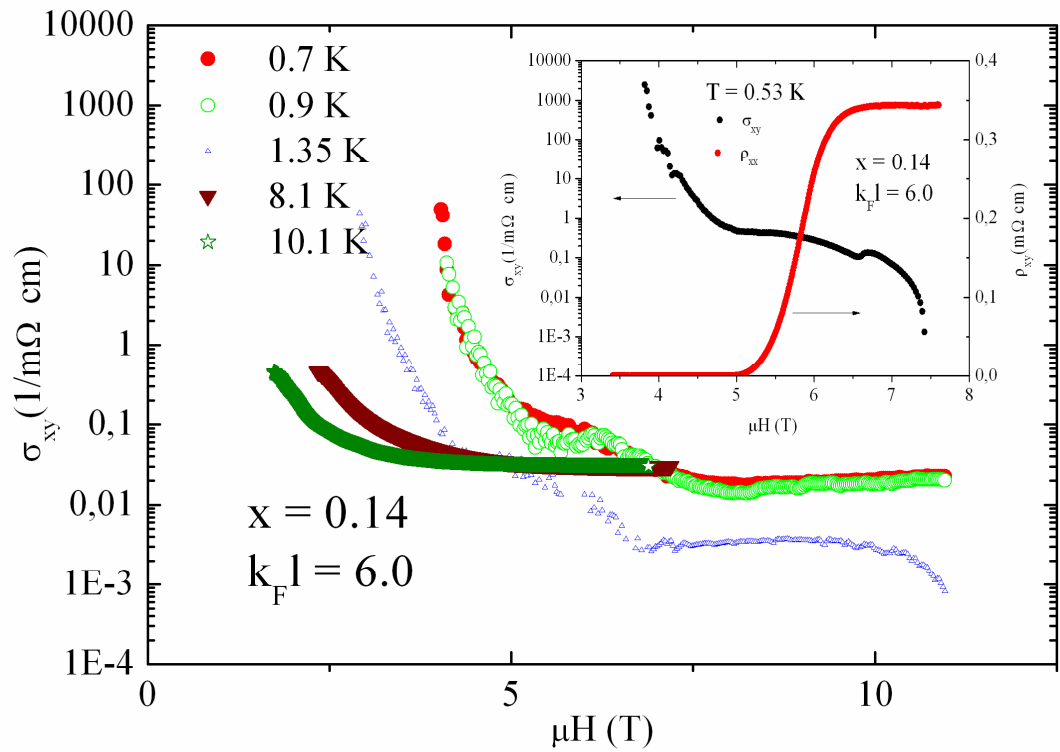